\documentclass{PoS}
\usepackage{amsmath}
\usepackage{amsfonts}
\usepackage{amssymb}

\newcommand{\Op}{\mathcal{O}} 

\title{A worm-inspired algorithm for the simulation of Abelian gauge theories}

\ShortTitle{A worm algorithm for gauge theories}

\author{\speaker{Tomasz Korzec}\\
        Institut f\"ur Physik, Humboldt Universit\"at zu Berlin, Newtonstrasse 15, 12489 Berlin, Germany\\
        E-mail: \email{korzec@physik.hu-berlin.de}}

\author{Ulli Wolff\\
        Institut f\"ur Physik, Humboldt Universit\"at zu Berlin, Newtonstrasse 15, 12489 Berlin, Germany\\
        E-mail: \email{uwolff@physik.hu-berlin.de}}

\abstract{We present an algorithm in which the all-order strong coupling expansion of
the Abelian U(1) gauge theory with Wilson plaquette action is sampled. In
addition to the vacuum closed surface graphs of the partition function we
propose to also allow for a class of defects (boundaries) related to Wilson
loops in the ensemble. The efficiency of our scheme in estimating various
observables is compared to a standard Metropolis algorithm.
\begin{flushright} HU-EP-10/67 \end{flushright}
\begin{flushright} SFB/CCP-10-94 \end{flushright}
}

\FullConference{The XXVIII International Symposium on Lattice Field Theory, Lattice2010\\
		June 14-19, 2010\\
		Villasimius, Italy}

\begin{document}

\section{Worm algorithms}
Monte-Carlo simulation algorithms based on all-order strong coupling or hopping parameter expansions
have been known for a long time~\cite{Berg:1981jy}. Yet only the relatively recent ideas of Prokof'ev and
Svistunov~\cite{Prokof'ev:2001zz} opened the way
to the formulation of highly efficient algorithms. Instead of sampling the partition function of a model
the new class of algorithms samples an enlarged ensemble that also contains the two-point-functions 
at all possible separations.
The ``worm algorithms'' that can be formulated in the loop gas representation of such an
ensemble typically show hardly any critical slowing down. In addition the non-standard formulation often makes
improved estimators for key observables possible. The basic idea behind worm algorithms is very general
and can be applied to a large class of statistical models. Efficient algorithms have been developed
for the one (real or complex) component $\phi^4$ theory and tested in its Gaussian and 
Ising or XY-model limits as well as at some intermediate values of the
coupling~\cite{Prokof'ev:2001zz,Deng:2007jq,Wolff:2008km,Wolff:2009ke,Janke:2009rb,Vierhaus:2010}.
More complicated systems like nonlinear $O(N)-\sigma$ models~\cite{Wolff:2009kp}, $CP(N-1)$
models~\cite{Wolff:2010qz} or two dimensional fermionic systems~\cite{Wolff:2008xa, Wenger:2008tq}
have been successfully simulated as well.
Some of the recent development has been summarized during this conference~\cite{Wolff:2010lat}.

In the following we make a first attempt to generalize the idea behind worm algorithms to the case of gauge theories. We formulate an algorithm that samples the (generalized) partition function of an Abelian $U(1)$
gauge theory.

\section{Strong coupling expansion}
We start from a Wilson plaquette gauge action
\begin{equation}\label{PlaquetteAction}
   S[U] = -\beta \sum_{x,\mu<\nu} {\rm Re}\left[U(x,\mu)U(x+\hat\mu, \nu)U^{-1}(x+\hat\nu,\mu)U^{-1}(x,\nu) \right]\, ,
\end{equation}
where $x$ is a site on a $D$-dimensional hypercubic periodic lattice of extent $L_\mu$ in direction
$\mu =1\ldots D$ and $U(x,\mu) \in U(1)$ denotes the gauge field on the link that connects site $x$ with site $x+\hat\mu$. We use lattice units throughout.

Observables in this model are products of link variables
\begin{equation}
   U^j = \prod_{x\mu} U(x,\mu)^{j(x,\mu)}\, ,
\end{equation}
where we have introduced an integer valued external field $j(x,\mu) \equiv j_\mu(x) \in {\mathbb Z}$. One is
interested in their expectation values
\begin{eqnarray}
   \langle U^j \rangle = \frac{Z[j]}{Z[0]}\, , \\
   Z[j] = \int DU\ e^{-S[U]} U^j \, ,
\end{eqnarray}
where $DU$ denotes the invariant measure on $U(1)$ on all links.
Gauge and center invariance of the action causes expectation values to vanish unless the external field 
satisfies
\begin{eqnarray}
   \partial^*_\mu j_\mu(x) &=& 0\, , \qquad {\rm and} \\
   \sum_x j_\mu(x)         &=& 0 \, ,
\end{eqnarray}
where $\partial^*$ denotes the backward nearest neighbor difference.

On each plaquette one can expand the exponential
\begin{equation}
   e^{\beta {\rm Re}(U)} = \sum_{m,n = 0}^{\infty} \left(\frac{\beta}{2}\right)^{m+n} \frac{U^{n-m}}{m!n!} 
                         = \sum_{n=-\infty}^{+\infty} I_n(\beta) U^n \, .
\end{equation}
The summation variables $n(x; \mu,\nu) \equiv n_{\mu\nu}(x)$ can be defined to be antisymmetric in $\mu$,$\nu$.
At this stage the original group integrals can be carried out leaving behind constraints. The partition 
function takes the form
\begin{equation}\label{eq_Zjexpanded}
   Z[j] = \sum_{\{n\}} \left(\prod_{x,\mu<\nu} I_{n_{\mu\nu}(x)}(\beta) \right)\delta[\partial^* n - j] \, ,
\end{equation}
with the abbreviation for the constraints
\begin{equation}
   \delta[\partial^* n - j] = \prod_{x\mu} \delta_{\partial_\nu^* n_{\nu\mu}(x), j_\mu(x)} \, .
\end{equation}
The representation eq.~(\ref{eq_Zjexpanded}) has been used as a starting point for Monte-Carlo
simulations in \cite{Sterling:1983fs} and more recently in \cite{Azcoiti:2009md}. The authors 
of~\cite{Azcoiti:2009md} find that a variant of
their algorithm, that is based on an expansion of $Z[0]$ of the $Z_2$ gauge theory only, 
suffers from critical slowing down with a dynamical exponent similar to that of a local algorithm in the
standard formulation of the model.

Inspired by worm-algorithms we pursue a different direction. In spin systems the crucial
algorithmic idea was to consider an enlarged system that allows for defects related to
two-point functions. Similarly in our case we consider the enlarged ensemble
\begin{equation}
   {\mathcal Z} = \sum_{\{j\}} \rho^{-1}[j]\ Z[j] \, .
\end{equation}
The sum is over all possible external fields $j_\mu(x)$ and, as in~\cite{Wolff:2008km}, each contribution is
weighted by some non-negative weight $\rho$. $\mathcal Z$ contains now also graphs with boundaries
rather than closed surfaces only. The partition function can be written in its
final form
\begin{equation}\label{eq_enlargedEnsemble}
   {\mathcal Z} = \sum_{\{n\}} \left(\prod_{x,\mu<\nu} I_{n_{\mu\nu}(x)}(\beta)\right)  \rho^{-1}[\partial^* n]\, .
\end{equation}
Expectation values with respect to this ensemble are formed in the usual way
\begin{equation}
   \langle\langle \Op[n] \rangle\rangle = \frac{1}{\mathcal Z} \sum_{\{n\}} \Op[n] \left(\prod_{x,\mu<\nu} I_{n_{\mu\nu}(x)}(\beta)\right)  \rho^{-1}[\partial^* n]\, ,
\end{equation}
and for later convenience and in analogy to ref.~\cite{Wolff:2009kp} we also define
an expectation value with respect to the vacuum ensemble by
\begin{equation}
   \langle\langle\Op[n]\rangle\rangle_0 = \frac{\langle\langle \Op[n] \delta[\partial^*n]\rangle\rangle}{\langle\langle\delta[\partial^*n]\rangle\rangle}\, .
\end{equation}

One can relate observables in the enlarged ensemble to those in the original one.
One way to measure the expectation value of an arbitrary non-vanishing correlator, 
given by some $j_\mu(x)$, is to find some background plaquette-field $k_{\mu\nu}(x)$ that 
solves the constraint $\partial^* k = j$. The observable
\begin{equation}\label{eq_generalObs}
   \Op[n] = \prod_{x,\mu<\nu} \frac{I_{n_{\mu\nu}(x)+k_{\mu\nu}(x)}(\beta)}{I_{n_{\mu\nu}(x)}(\beta)}
\end{equation}
is then an estimator for the desired correlator
\begin{equation}
   \langle U^j \rangle = \langle\langle \Op[n] \rangle\rangle_0 \, .
\end{equation}
In practice $k$ will be zero almost everywhere. For rectangular loops the minimal number of
factors differing from one necessary in the product eq.~(\ref{eq_generalObs}) equals their area.

\section{Monte-Carlo updates}
We use the freedom in choosing the weight function $\rho^{-1}$ to restrict the possible
external fields or boundaries to the subset containing only one non-intersecting
loop with zero winding
number with respect to all torus directions. Such a loop can be completely characterized by the
ordered cyclic set of sites through which it passes
\begin{equation}
   {\mathcal W} = \{x_1, x_2, \ldots , x_{P({\mathcal W})}\}, \quad x_{P({\mathcal W})+1} = x_1\, .
\end{equation}
Consecutive sites are nearest neighbors $x_{i+1} = x_i + \hat\mu_i$. If we extend
the possible unit vectors to include negative directions: $\widehat{-\mu} \equiv -\hat\mu$,
plaquettes can be labeled in several equivalent ways
\begin{equation}
   n_{\mu\nu}(x) = -n_{-\mu\nu}(x+\hat\mu) = -n_{\mu-\nu}(x+\hat\nu) = n_{-\mu-\nu}(x+\hat\mu+\hat\nu) \, .
\end{equation}
To stochastically sample the sum eq.~(\ref{eq_enlargedEnsemble}) we perform a sequence of
local updates of which each by itself satisfies detailed balance. The updates consist of a
proposal ${\mathcal W}\to {\mathcal W}'$ followed by a Metropolis accept/reject step.
\subsection{Flips}
\begin{figure}[ht]
   \centering
   \includegraphics[height=2.5cm]{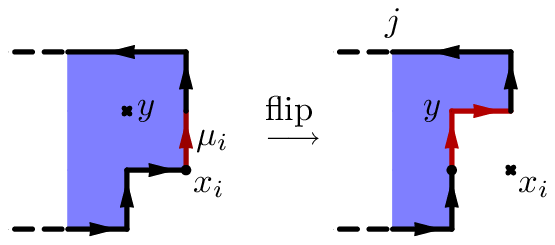}\\
   \includegraphics[height=2.5cm]{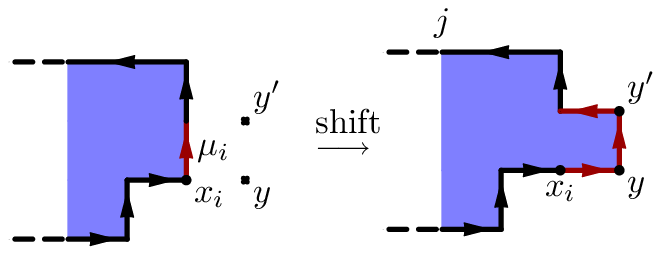}
\caption{Flip (upper part) and shift (lower part) updates. The plaquette field in the shaded area differs 
by one unit with respect to the non-shaded areas.}\label{fig_updates}
\end{figure}

If the perimeter $P({\mathcal W})$ is larger than four we pick a random site $x_i$ on the loop and
form $y = x_{i+1} - x_i + x_{i-1}$. If $y\in {\mathcal W}$ the old configuration is kept, 
otherwise the proposal 
is to form  $\mathcal W'$ by replacing $x_i \to y$
and to adjust the plaquette field accordingly
\begin{equation}
   n_{-\mu_{i-1}\mu_i}(x_i) \to n_{-\mu_{i-1}\mu_i}(x_i) + 1 \, .
\end{equation}
Such a proposal is accepted with probability $min(1,q_{\rm flip})$ with
\begin{equation}
   q_{\rm flip} = \frac{I_{n_{-\mu_{i-1}\mu_i}(x_i) + 1}(\beta)}{I_{n_{-\mu_{i-1}\mu_i}(x_i)}(\beta)} \ \frac{\rho({\mathcal W})}{\rho({\mathcal W'})} \, .
\end{equation}
This update changes the shape of the loop without altering its perimeter. It is displayed in fig.~\ref{fig_updates}.

\subsection{Shifts}
A site $x_i$ and a direction $\hat\nu$ orthogonal to $\hat\mu_i$ is chosen randomly. Two sites,
$y=x_i+\hat\mu$ and $y'=x_{i+1}+\hat\nu$ are constructed. A proposal is made in either of two cases
\begin{enumerate}
 \item If neither $y \in {\mathcal W}$ nor $y' \in {\mathcal W}$, it is proposed to extend the loop
 by two sites
 \begin{equation}
    {\mathcal W'} = \{x_1,\ldots, x_i,y,y',x_{i+1},\ldots \} ,
 \end{equation}
 together with the necessary change in the plaquette field: $n_{\nu\mu_i}(x_i) \to n_{\nu\mu_i}(x_i)+1$\, .
 In analogy to the flips we accept with
 \begin{equation}
    q_{\rm shift} = \frac{I_{n_{\nu\mu_i}(x_i) + 1}(\beta)}{I_{n_{\nu\mu_i}(x_i)}(\beta)} \ \frac{\rho({\mathcal W})}{\rho({\mathcal W'})} \, .
 \end{equation}

 \item If $y=x_{i-1}$ and $y'=x_{i+2}$ the reverse move is proposed, i.e. $x_i$ and $x_{i+1}$ are proposed
 to be removed from the loop. The acceptance probability is again $min(1,q_{\rm shift})$.
\end{enumerate}
Otherwise no move is made.

\subsection{Non-local updates}
With free boundary conditions the alternation of the two updates proposed so far 
would be sufficient to guarantee ergodicity.
On a torus however, there exist contributions to the partition function that
are not yet sampled. This is fixed by another update:\\
A plaquette $x,\mu, \nu\neq\mu$ is chosen randomly. The subsets of sites
which belong to the two dimensional plane through $x$ spanned by $\hat\mu$ and $\hat\nu$
is denoted by $\Gamma$. The proposal is to change the plaquette field according to\footnote{The inverse
move is realized by swapping $\mu\leftrightarrow\nu$.}
\begin{equation}
    {\rm for\ all}\ x\in\Gamma:\qquad n_{\mu\nu}(x) \to n_{\mu\nu}(x) + 1
\end{equation}
and it is accepted with 
\begin{equation}
   q_{\rm plane} = \prod_{x\in\Gamma} \frac{I_{n_{\mu\nu}(x)+1}(\beta)}{I_{n_{\mu\nu}(x)}(\beta)}\, .
\end{equation}
The acceptance rate for this proposal is tiny unless the volume is small.

The last update that we have implemented is proposed whenever $\mathcal W$ is planar,
or alternatively whenever it has exactly 4 corners. The proposal is to shift the whole
loop by one unit in a direction perpendicular to its plane. This update can be built up from the 
elementary shifts and flips, but it turns out to reduce autocorrelation times considerably if it is
introduced as a separate update step.

One iteration of our algorithm consists of order of volume many shift, flip and planar-loop-shift
updates followed by one plane update. The costs are comparable to those of a Metropolis sweep
in the standard formulation, i.e. O(volume).

\section{Performance of the algorithm}
\begin{figure}[ht]
 \centering
 \includegraphics[width=0.43\linewidth]{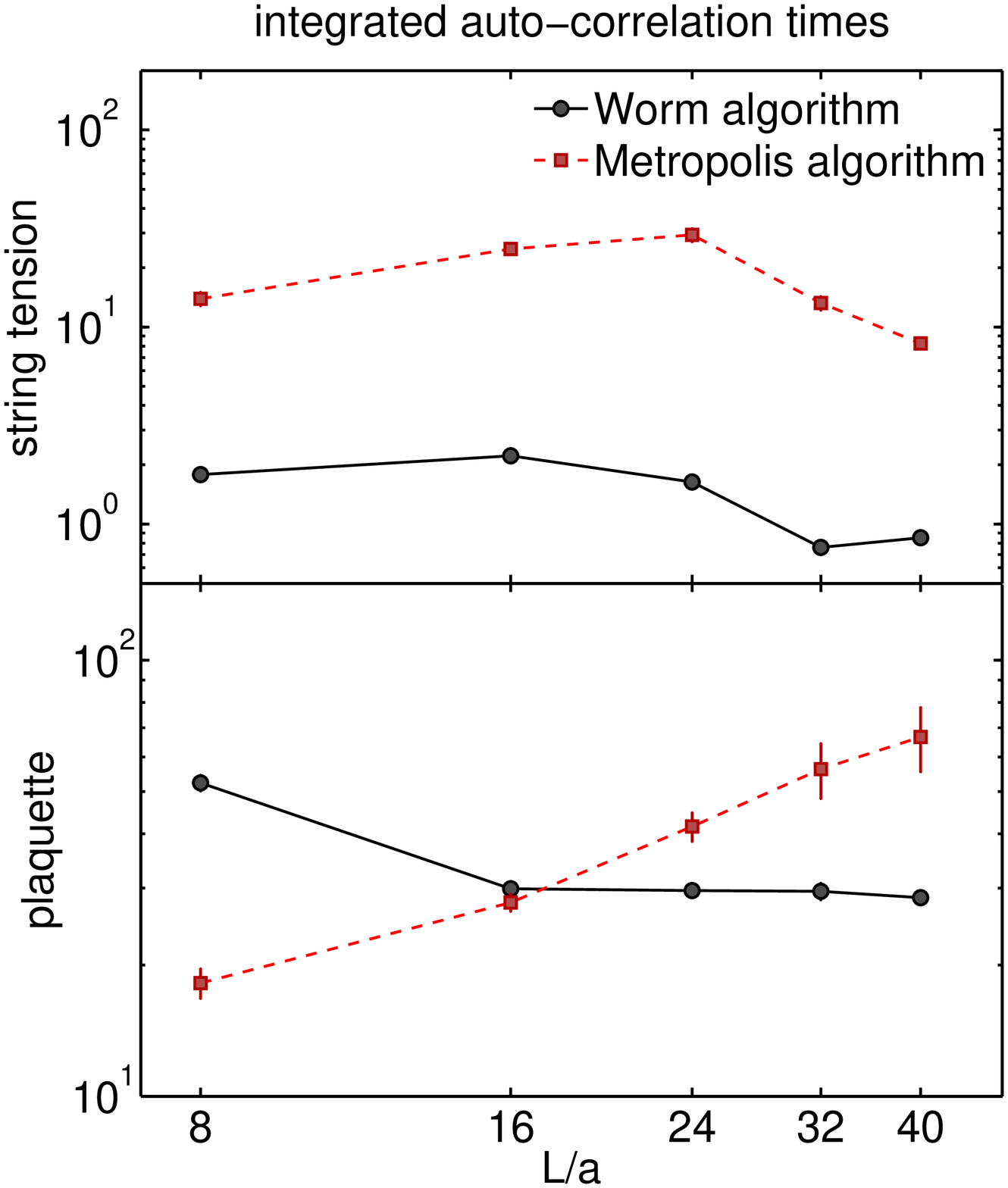}\hspace{0.1\linewidth}
 \includegraphics[width=0.43\linewidth]{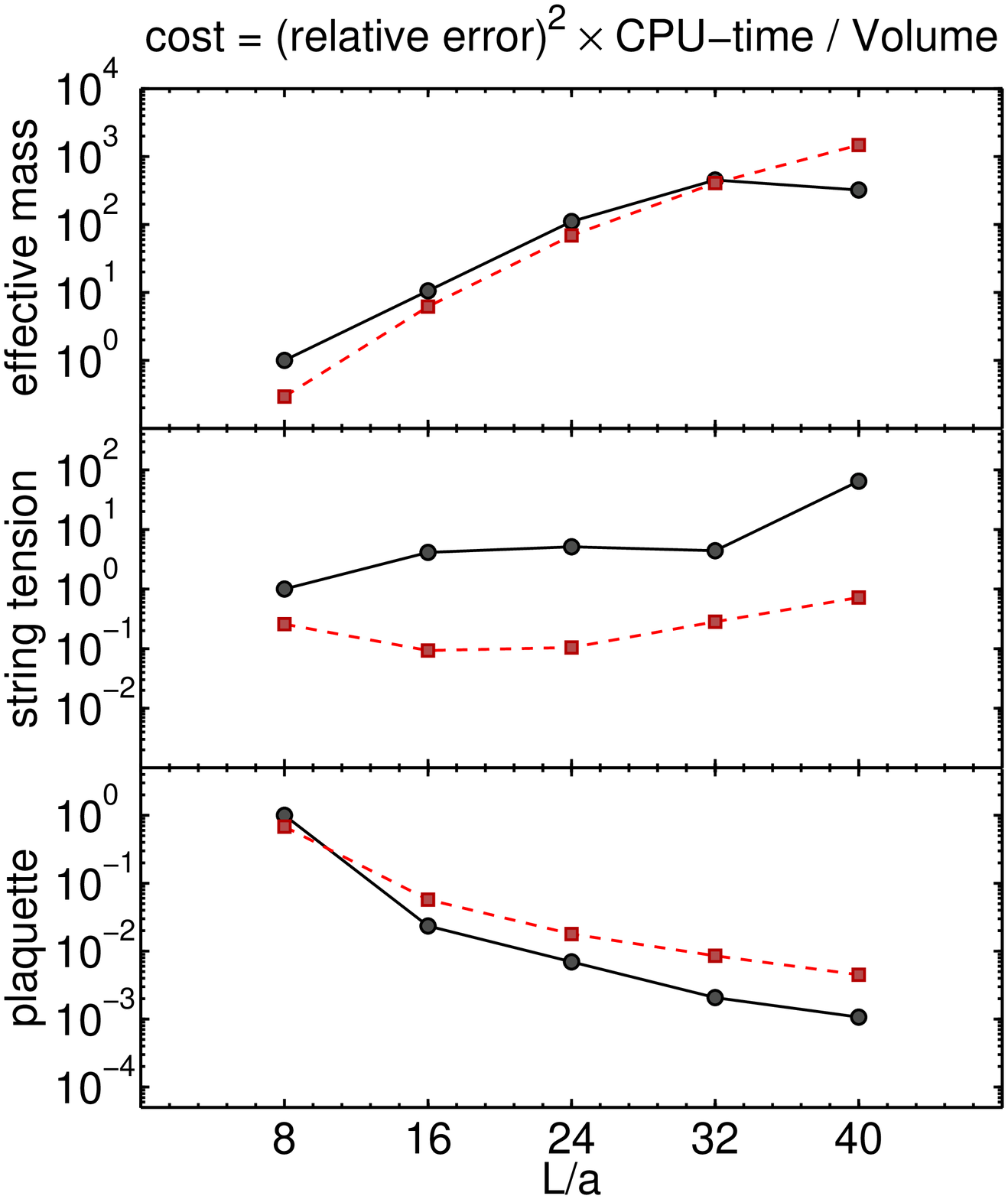}
 \caption{A comparison of the new algorithm with a standard Metropolis one. The left 
 panel shows integrated autocorrelation times of two observables in units of sweeps or iterations. 
 The right panel compares the costs. The autocorrelation times of the effective mass are consistent with
 0.5 for this set of lattices and are not plotted.}\label{fig_results}
\end{figure}

Abelian gauge theory is particularly well understood in three dimensions. For the model with
Villain action, which is believed to lie in the same universality class, the existence of a mass gap $m$ and confinement at every value of $\beta$ have been 
proven~\cite{Gopfert:1981er}. The continuum limit at fixed $m$ is believed to describe free massive bosons.
A recent numerical work with the Wilson action~\cite{Loan:2002ej} showed that the mass gap is indeed well
described by
\begin{equation}\label{eq_massgap}
   mL = 5.23(11)\frac{L}{a}\sqrt{8\pi^2\beta}\, \exp\left[-0.2527 \pi^2\beta \right] \, .
\end{equation}
To test our algorithm we use eq.~(\ref{eq_massgap}) to keep the physical volume constant at $mL\approx 6$ while increasing
the resolution $L/a\in\{8,16,24,32,40 \}$. As weight function we take an exponential that depends on the
loop perimeter $\rho[{\mathcal W}] = \exp(\theta(P[{\mathcal W}]-2))$, and tune the ``loop tension'' $\theta$ to
an algorithmically close to optimal value. More precisely, the cost to estimate a $L/4 \times L/4$ Wilson loop to a given precision is minimalized.
We measure expectation values of rectangular Wilson loops as well as of plaquette-plaquette correlators.
The autocorrelation times and a cost-indicator (CPU-time $\times$ relative error squared divided by volume) 
of the average plaquette,
the string tension
and an effective mass from the plaquette-plaquette correlator at separation $L/4$ are monitored.
The same calculations are repeated with a standard Metropolis algorithm, in which the proposals are to
change the gauge field on a link $e^{i\phi} \to e^{i(\phi+\Delta\phi)}$. In this case the number of
sweeps per measurement as well as the size of the interval from which $\Delta \phi$ is drawn
are kept at their optimal values. The optimal number of sweeps per measurement 
is determined only on the $16^3$ lattice
and kept at the same value (i.e. 15) for the others. The size of the interval is determined on each lattice
separately. Expectation values obtained with both algorithms are consistent with each other and
a comparison of autocorrelation times and costs is shown in Fig.~\ref{fig_results}.

\section{Conclusions}
We have extended the concept of worm algorithms to Abelian gauge theories.
Instead of a loop gas we deal with a surface ensemble. The worm is replaced by an
open surface and a Wilson loop plays the role analogous to the worm's head and tail. All
standard observables can be estimated in this model. In first numerical tests in three dimensions
no significant critical slowing down could be observed, but larger correlation lengths
will be necessary to make a definite statement. On the presently available lattices also a standard
Metropolis algorithm performs relatively well.

Abelian gauge theories with different actions, like the Villain model or Wegner's $Z_2$ model can
presumably be treated in the same way. Whether matter fields can be incorporated
into the algorithm or whether an extension to non Abelian models is feasible remains
to be investigated.

\bibliographystyle{JHEP-2}
\bibliography{proc}

\providecommand{\href}[2]{#2}\begingroup\raggedright\begin{thebibliography}{10}

\bibitem{Berg:1981jy}
B.~Berg and D.~F{\"o}rster {\em Phys. Lett.} {\bf B106} (1981) 323.

\bibitem{Prokof'ev:2001zz}
N.~Prokof'ev and B.~Svistunov {\em Phys. Rev. Lett.} {\bf 87} (2001) 160601.

\bibitem{Deng:2007jq}
Y.~Deng, T.~M. Garoni and A.~D. Sokal {\em Phys. Rev. Lett.} {\bf 99} (2007)
  110601 [\href{http://arXiv.org/abs/cond-mat/0703787}{{\tt
  arXiv:cond-mat/0703787}}].

\bibitem{Wolff:2008km}
U.~Wolff {\em Nucl. Phys.} {\bf B810} (2009) 491--502
  [\href{http://arXiv.org/abs/0808.3934}{{\tt arXiv:0808.3934}}].

\bibitem{Wolff:2009ke}
U.~Wolff {\em Phys. Rev.} {\bf D79} (2009) 105002
  [\href{http://arXiv.org/abs/0902.3100}{{\tt arXiv:0902.3100}}].

\bibitem{Janke:2009rb}
W.~Janke, T.~Neuhaus and A.~M.~J. Schakel {\em Nucl. Phys.} {\bf B829} (2010)
  573--599 [\href{http://arXiv.org/abs/0910.5231}{{\tt arXiv:0910.5231}}].

\bibitem{Vierhaus:2010}
I.~Vierhaus, {\it {S}imulation of {$\phi^4$} {T}heory in the {S}trong
  {C}oupling {E}xpansion beyond the {I}sing {L}imit},  Diploma thesis, Humboldt
  Universit{\"a}t zu Berlin, 2010, and\\
  T.~Korzec, I.~Vierhaus, U.~Wolff, in preparation.

\bibitem{Wolff:2009kp}
U.~Wolff {\em Nucl. Phys.} {\bf B824} (2010) 254--272
  [\href{http://arXiv.org/abs/0908.0284}{{\tt arXiv:0908.0284}}].

\bibitem{Wolff:2010qz}
U.~Wolff {\em Nucl. Phys.} {\bf B832} (2010) 520--537
  [\href{http://arXiv.org/abs/1001.2231}{{\tt arXiv:1001.2231}}].

\bibitem{Wolff:2008xa}
U.~Wolff {\em Nucl. Phys.} {\bf B814} (2009) 549--572
  [\href{http://arXiv.org/abs/0812.0677}{{\tt arXiv:0812.0677}}].

\bibitem{Wenger:2008tq}
U.~Wenger {\em Phys. Rev.} {\bf D80} (2009) 071503
  [\href{http://arXiv.org/abs/0812.3565}{{\tt arXiv:0812.3565}}].

\bibitem{Wolff:2010lat}
U.~Wolff \pos{PoS(Lattice 2010)020} (2010)
  [\href{http://arXiv.org/abs/1009.0657}{{\tt arXiv:1009.0657}}].

\bibitem{Sterling:1983fs}
T.~Sterling and J.~Greensite {\em Nucl. Phys.} {\bf B220} (1983) 327.

\bibitem{Azcoiti:2009md}
V.~Azcoiti, E.~Follana, A.~Vaquero and G.~Di~Carlo {\em JHEP} {\bf 08} (2009)
  008 [\href{http://arXiv.org/abs/0905.0639}{{\tt arXiv:0905.0639}}].

\bibitem{Gopfert:1981er}
M.~G{\"o}pfert and G.~Mack {\em Commun. Math. Phys.} {\bf 82} (1981) 545.

\bibitem{Loan:2002ej}
M.~Loan, M.~Brunner, C.~Sloggett and C.~Hamer {\em Phys. Rev.} {\bf D68} (2003)
  034504 [\href{http://arXiv.org/abs/hep-lat/0209159}{{\tt
  arXiv:hep-lat/0209159}}].

\end{thebibliography}\endgroup

\end{document}